\documentclass[twocolumn,showpacs,preprintnumbers,amsmath,amssymb]{revtex4}
\usepackage{graphicx}
\usepackage{epsfig}
\usepackage{dcolumn}
\usepackage{bm}
\begin{document}
\preprint{} 
\title{Effect of Delocalized Vortex Core States on the Specific Heat of Nb}
\author{J.E.~Sonier$^{1,2}$, M.F.~Hundley$^3$ and J.D.~Thompson$^3$}
\affiliation{$^1$Department of Physics, Simon Fraser University, Burnaby, 
British Columbia V5A 1S6, Canada\\
$^2$Canadian Institute for Advanced Research, Toronto, Ontario M5G 1Z8, Canada \\
$^3$Los Alamos National Laboratory,
Los Alamos, New Mexico 87545, USA}
\date{\today} 
\begin{abstract} 
The magnetic field ($B$) dependence of the electronic specific heat for a
simple BCS type-II superconductor has been determined from 
measurements on pure niobium (Nb). Contrary to expectations, 
the electronic specific heat coefficient $\gamma(T, B)$ is observed to be a 
sublinear function of $B$ at fields above the lower critical field $H_{c1}$. 
This behavior is attributed to the delocalization of quasiparticles bound to 
the vortex cores. The results underscore the ambiguity of interpretation that 
arises in specific heat studies of this kind on newly discovered type-II 
superconductors, and also emphasize the need to do such measurements under  
field-cooled conditions. 
\end{abstract} 
\maketitle 
Specific heat measurements on type-II superconductors in 
the vortex state are sensitive to low-energy quasiparticle excitations. 
In fully-gapped superconductors the low-lying excitations 
are traditionally thought of as being confined to the vortex cores \cite{Caroli:64}. In this 
case the coefficient of the linear term in the specific heat $c$ as a function of 
temperature, {\it i.e.} the `Sommerfeld coefficient' $\gamma$, 
is proportional to the density of vortices, and hence a linear function
of the internal magnetic field $B$ \cite{Fetter:69}. 
The situation is different in superconductors with a
highly anisotropic energy gap or gap nodes. In these cases there is a major contribution to 
the electronic specific heat at low temperatures from delocalized quasiparticles. 
The flow of superfluid around a vortex lowers the energy of quasiparticles delocalized
near the gap minima, resulting in a finite density of states (DOS) at the Fermi
energy and a corresponding nonlinear field dependence for $\gamma(B)$ \cite{Volovik:93}.
For an anisotropic gap, such behavior is expected only if the energy shift exceeds
the gap minimum. Thus, $\gamma(B)$ tells us something
about the gap structure. 

What has not been considered in the interpretation of many such 
studies, is that even for the 
case of a simple BCS superconductor characterized by an isotropic energy gap, 
the quasiparticles are not strictly confined to single vortices.
Quasiparticle states bound to the vortex cores become
delocalized when their spatially extended wave functions overlap those from
nearest-neighbor vortices. The overlap is enhanced at higher magnetic fields 
where the vortices are closer together, but there is also some overlap as soon as 
there is more than one vortex in the sample. 
Likewise, the degree of delocalization
increases with increasing temperature due to thermal population of
higher-energy core states, characterized by wave functions
extending further out from the core center. 
This is the so-called `Kramer-Pesch (KP) effect' \cite{Kramer:74}. 

Recently, the field dependence of the vortex electronic excitations
has been confirmed by thermal conductivity \cite{Boaknin:03} and muon spin rotation 
($\mu$SR) \cite{Sonier:04}
experiments on V$_3$Si. By detecting the change in the vortex core size, $\mu$SR 
measurements on V$_3$Si have confirmed that the delocalized quasiparticles 
detected by thermal conductivity originate from the vortex cores. These delocalized
core states should have a profound influence on the field 
dependence of the specific heat, which probes the electronic DOS
at the Fermi level, both inside and outside of the vortex cores. 
Recent calculations \cite{Nakai:04}
in the framework of the quasi-classical Eilenberger theory show that for an isotropic 
energy gap, $\gamma(B) \! \propto \! B$ behavior persists only up to a
crossover field $B^*$, above which the overlap of core states results in a 
sublinear dependence of $\gamma(B)$ on $B$. Strictly speaking $B^*$ is 
temperature dependent, reduced by thermal population 
of the more energetic and spatially extended core states. 
  
To date there has been no clear experimental verification of a 
crossover field $B^*(T)$ in a fully-gapped type-II superconductor. 
While there have been numerous specific heat studies on BCS type-II 
superconductors, rarely is $\gamma(B)$ observed to be a linear function 
of $B$. One reason is there are very few ``simple'' BCS type-II 
superconductors. For example, specific heat measurements 
on the widely studied superconductor 2$H$-NbSe$_2$ show that
$\gamma(B)$ is a nonlinear function of $B$ even at low fields 
\cite{Sanchez:95,Sonier:99,Hanaguri:03}.  
There is now good evidence that 2$H$-NbSe$_2$ is a multiband superconductor, with
a different size energy gap on two different sheets of the Fermi surface 
\cite{Boaknin:03,Yokoya:01}. The nonlinearity of $\gamma(B)$ can be
attributed to two-band superconductivity \cite{Nakai:02}, giving rise to
a large vortex-core size at low fields \cite{Sonier:99,Callaghan:05}. 

Even the cubic A-15 compound V$_3$Si is not so simple. For a magnetic field applied 
along the $\langle 001 \rangle$ direction, 
there is a substantial range of fields over which the hexagonal
vortex lattice either gradually distorts into a square lattice \cite{Sosolik:03,Sonier:04} or 
coexists with a square lattice \cite{Yethiraj:99}. This vortex lattice transition is accompanied
by delocalization of quasiparticle core states \cite{Boaknin:03,Sonier:04}, and occurs
at very low fields. Consequently, the specific heat is a linear function 
of $B$ only over a very narrow field range far below the upper critical 
field $H_{c2}$ \cite{Sonier:04b}.

\begin{figure}
\centerline{\epsfxsize=3.5in\epsfbox{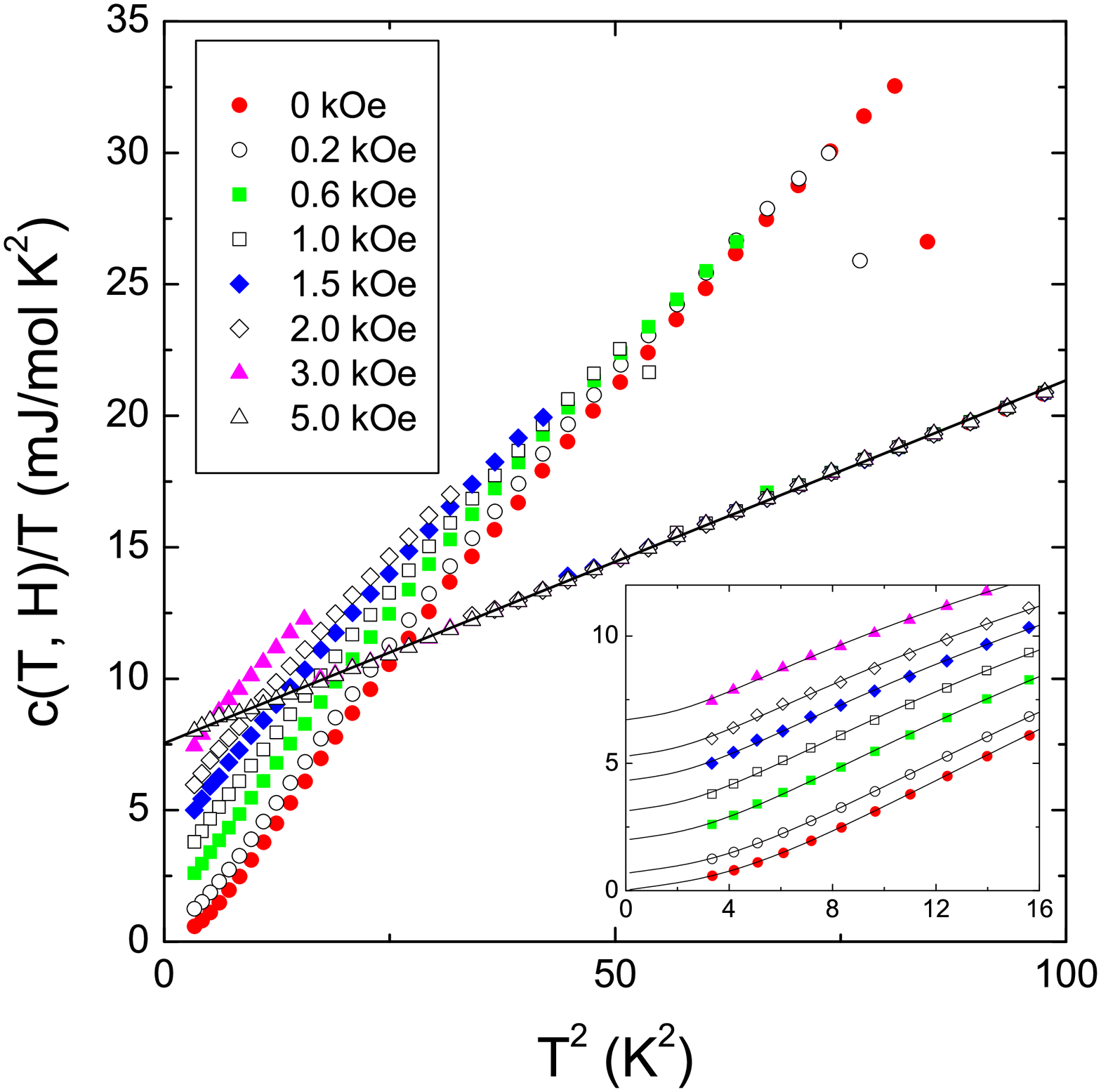}}
\caption{Temperature dependence of the specific heat at several different applied 
magnetic fields, plotted as $c(T, H)/T$ vs $T^2$. The measurements were taken
under FC conditions. The solid line through the normal state data at $H \! = \! 5$~kOe 
is a fit described in the main text. Inset: Extrapolation of the low-temperature data to 
0~K.}
\end{figure}

In view of these complications, here we investigate the field 
dependence of  $\gamma(B)$ in the elemental type-II superconductor niobium (Nb). 
A $\langle 001 \rangle$-oriented, 99.99~\% purity Nb single crystal was obtained from 
Goodfellow Corporation. The sample was in the shape of a short cylinder, $\sim \! 1.8$~mm
long with a diameter of $\sim \! 2$~mm. Magnetization measurements were made with
a SQUID magnetometer and specific heat measurements were made
using a Quantum Design Physical Properties Measurement System (PPMS) that
utilizes a thermal relaxation calorimeter. The calorimeter is composed of a sapphire
sample holder and the addenda heat capacity is essentially field-independent for the
fields employed in this work.
The external field $H$ was directed along [111], which under field-cooled 
(FC) conditions produces a well-ordered hexagonal vortex lattice in Nb for all 
fields $H_{c1} \! < \! H \! < \! H_{c2}$ \cite{Forgan:02}.
FC and zero-field cooled (ZFC) swept-field measurements of the 
sample magnetization $M(T, H)$ and the specific heat $c(T, H)$
were found to be greatly hampered by flux jumps.
Consequently, the FC measurements reported here were done    
by field cooling the sample from above $T_c$ to the desired temperature for each 
value of $H$. On the other hand, the ZFC measurements were done by first cooling
the sample in zero field and then increasing $H$. 
From ZFC measurements of 
$M(T)$ at 5~Oe, the Nb crystal was determined to have 
a sharp superconducting transition temperature of $T_c \! = \! 9.2$~K.  
The zero-temperature extrapolated value of the upper critical field
was determined to be $H_{c2}(0) \! \approx \! 4.5$~kOe, from
which the superconducting coherence length is calculated to be
$\xi \! = \! 270$~\AA. Just above $T_c$ the resistivity of the sample saturates 
with a value of 0.32~$\mu \Omega$-cm, from which we calculate the electron
mean free path to be $l \! = \! 2300$~\AA. Thus, our sample is well
within the clean limit.  

Measurements of the sample magnetization $M(H)$ were done to determine
the internal magnetic field
\begin{equation}
B = H -4 \pi M (1-N) \, ,
\end{equation}
where $N \! = \! 0.379$ is the demagnetization factor of our Nb crystal.
This value of $N$ was determined by ZFC measurements
of $M(H)$ at several temperatures in the Meissner phase, 
assuming full shielding ($B \! = \! 0$). Under FC conditions
the quantity $4 \pi M$ was found to be very small (with a maximum value of 
31~G at $H \! = \! 750$~Oe and $T \! = \! 2.25$~K), 
indicating that very little flux is expelled. 
Since $B \! \approx \! H$, it is safe to interpret our measurements
of $\gamma(T, H)$ as $\gamma(T, B)$. We note that this
equivalence is often assumed in specific heat studies without justification. 

\begin{figure}
\centerline{\epsfxsize=3.5in\epsfbox{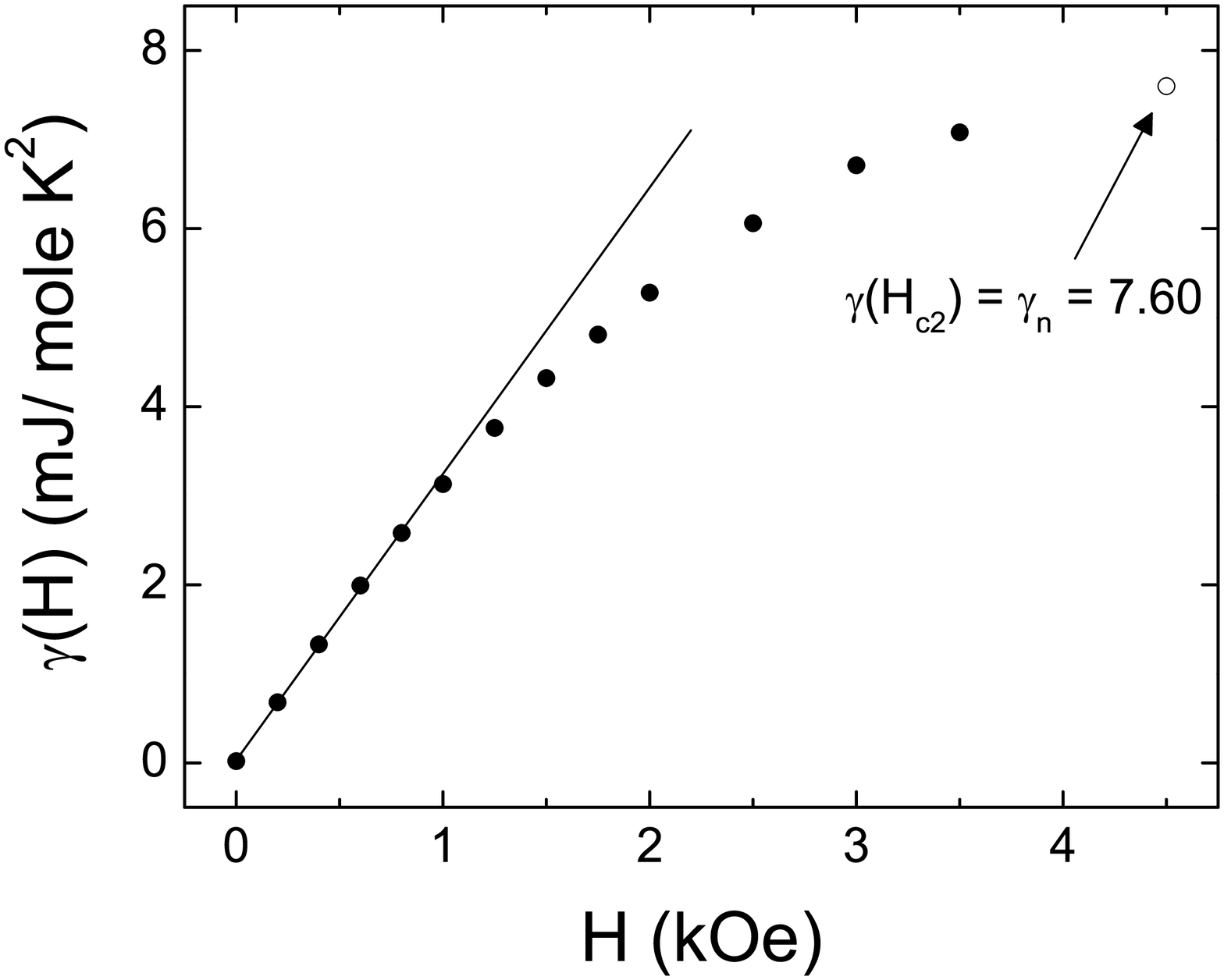}}
\caption{ Magnetic field dependence of the Sommerfeld coefficient. The open circle
indicates $\gamma$ at $H_{c2}$}.
\end{figure}

The normal-state specific heat of Nb is described by the relation
\begin{equation}
c_{\rm n}(T) = \gamma_{\rm n} T + \beta T^3 \, ,
\label{eq:normalC}
\end{equation}
where $\gamma_{\rm n}$ is the Sommerfeld constant in the normal state,
and $\gamma_{\rm n} T$ and $\beta T^3$ 
are the electronic and phonon contributions, respectively.
FC measurements of the temperature dependence of the
specific heat of Nb are plotted in Fig.~1 as $c(T, H)/T$ against $T^2$. 
The solid line in the main panel is a fit of the $H \! = \! 5$~kOe data 
below 10~K to Eq.~(\ref{eq:normalC}), yielding
$\gamma_{\rm n} = \! 7.60 \! \pm \! 0.02$~mJ/mol~K$^2$ and 
$\beta \! = \! 0.14 \! \pm \! 0.01$~mJ/mol~K$^4$.    

At $T \! \ll \! T_c$ the specific heat is described by
\begin{eqnarray}    
c(T, H) & = & \gamma(T, H) T + \beta T^3 + c_{\rm es}(T, 0) \\
            & = & \gamma(T, H) T + c(T, 0) \, ,
\label{eq:heatH}
\end{eqnarray}
where $c_{\rm es}(T, 0)$ is the electronic specific heat in the 
superconducting medium surrounding the vortex cores.
 To determine the Sommerfeld coefficient 
$\gamma(H) \! \equiv \! {\rm lim}_{T \rightarrow 0} \, c(T, H)/T$,
the low-$T$ data shown in the inset of Fig.~1 were extrapolated to 
$T \rightarrow 0$~K assuming
\begin{equation}    
\frac{c(T, H)}{T} = \gamma(T, H) + \beta T^2 \, + \frac{a e^{-b T_c/T}}{T}.
\label{eq:extrap}
\end{equation}
The last term is the limiting low-temperature BCS expression for
$c_{\rm es}(T, 0)$ divided by $T$. At $H \! = \! 0$~kOe, the fit to
Eq.~(\ref{eq:extrap}) yields 
$c_{\rm es}(T, 0)/\gamma_n T_c  \! = \! 10.19 \exp(-1.64 T_c/T)$.
As shown in Fig.~2, $\gamma(B)$ is a linear function of $B$ only up
to $H \! \approx \! 0.22 H_{c2}$.

\begin{figure}
\centerline{\epsfxsize=4.1in\epsfbox{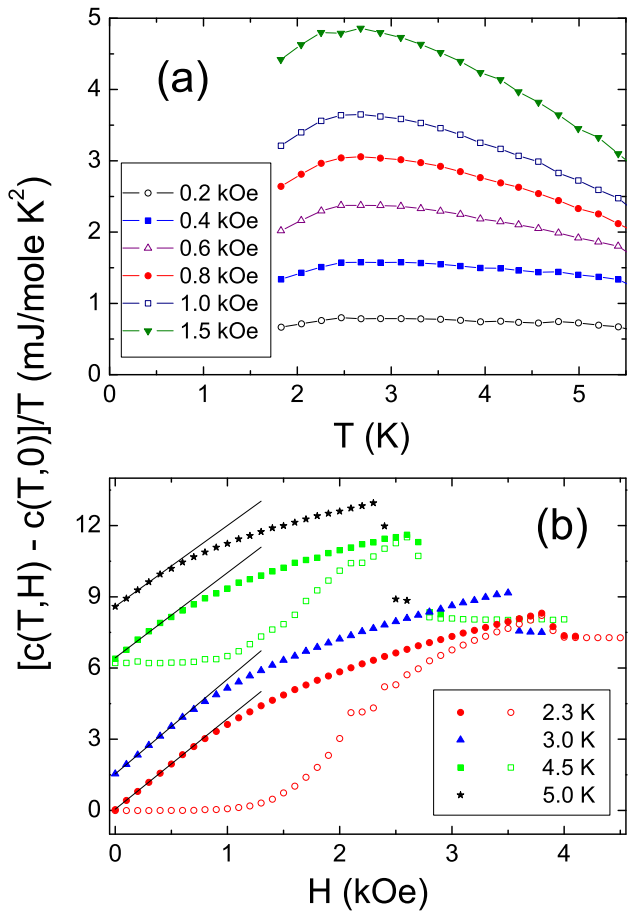}}
\caption{(a) Temperature dependence of $[c(T, H) \! - \! c(T, 0)]/T$ at several
fields measured under FC conditions. (b) Field dependence of $[c(T, H) \! - \! c(T, 0)]/T$ 
at several temperatures below $T_c$. The open and solid symbols correspond to ZFC 
and FC measurements, respectively. The downturn at high fields for each data set 
indicates $H_{c2}(T)$. For visual clarity the data at $T \! =$~3.0, 4.5 and 5.0~K 
are vertically offset.}
\end{figure}

The temperature and magnetic field dependences of the difference 
$[c(T, H) \! - \! c(T, 0)]/T$ are plotted in Fig.~3. At low temperatures
this quantity reflects $\gamma(T, H)$, as inferred from Eq.~(\ref{eq:heatH}). 
In Fig.~3(a) we see that below 2.5~K, 
$[c(T, H) \! - \! c(T, 0)]/T$ decreases with decreasing $T$. 
We attribute this behavior to shrinking of the vortex cores
({\it i.e} the KP effect), which reduces the contribution of the zero-energy 
DOS per vortex to the specific heat. The effect is
more pronounced at higher $H$, due to the increased density of vortices.
Above 2.5~K, $[c(T, H) \! - \! c(T, 0)]/T$ decreases with increasing $T$ due 
to the subtraction of the $H  \! = \! 0$ specific heat jump at $T_c$, and hence
does not reflect the temperature dependence of $\gamma$.
From the field dependence of $[c(T, H) \! - \! c(T, 0)]/T$ measured under 
FC conditions (see Fig.~3(b)), we conclude that $\gamma(B)  \! \propto \! B$
up to a temperature-dependent crossover field. 

\begin{figure}
\centerline{\epsfxsize=3.8in\epsfbox{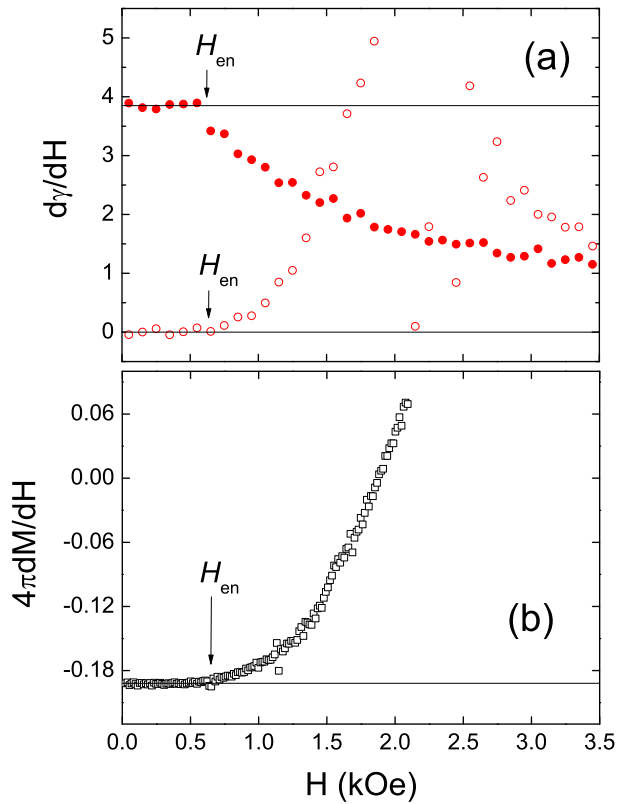}}
\caption{Magnetization and specific heat data at $T \! = \! 2.3$~K. 
(a)  FC (solid circles) and ZFC (open circles) measurements of $d \gamma/dH$,
and (b) ZFC measurements of $dM/dH$. The ZFC measurements were done
by first cooling the sample in zero field to $T \! = \! 2.3$~K and then increasing $H$.} 
\end{figure}

To clearly see the field range over which $\gamma$ is a linear function of $B$, 
in Fig.~4(a) we show the derivative $d \gamma /dH$ of the 
$\gamma \! \approx \! [c(T, H) \! - \! c(T, 0)]/T$ data at $T \! = \! 2.3$~K. 
For both FC and ZFC measurements, $d \gamma /dH$ 
is constant only at low fields. In this same low-field range, $\gamma(T, H)$ is
determined to be zero from the ZFC measurements. Thus, the nonzero
value of $d \gamma /dH$ in the FC data at low fields must result from 
trapped vortices below the lower critical field $H_{c1}$. To see that this is indeed
the case, we determined the field of first vortex entry $H_{\rm en}(T)$ from 
ZFC measurements of $M(H)$ for increasing $H$. 
As shown in Fig.~4(b), the increase of $dM/dH$ 
at $H_{\rm en}$ agrees with the decrease 
(increase) of $d \gamma /dH$ in the FC (ZFC) measurements of the specific heat. 
This was also determined to be the case at the other temperatures.
We note that $H_{c1}(T)$ is somewhat lower than $H_{\rm en}(T)$, if one assumes there is a 
Bean-Livingston barrier \cite{Bean:64} that both impedes the entry of flux when the field is 
increased above $H_{c1}(T)$, and traps flux below $H_{c1}(T)$ in the FC 
measurements.
  
The observation of the simple relation $\gamma(B) \! \propto \! B$ only in FC 
measurements below $H_{\rm en}(T)$, indicates that the trapped vortices form
a highly-disordered lattice, in which they behave as isolated vortices.
The disorder apparently disrupts the connection between the local DOS of
nearest-neighbor vortices that is found in a well-ordered lattice \cite{Ichioka:99}. 
At fields immediately above $H_{\rm en}(T)$, $\gamma(B)$ exhibits a sublinear 
dependence on $B$. As explained in Ref.~\cite{Ichioka:99}, the sublinear dependence 
of $\gamma(B)$ on $B$ is due to the shrinkage of the vortex cores that 
occurs as a result of the delocalization of the higher-energy core states.
While this may seem at odds with electronic thermal conductivity measurements of Nb
that indicate only a small degree of quasiparticle delocalization just above
$H_{c1}$ \cite{Lowell:70}, the details of the experimental method must be
considered. In Ref.~ \cite{Lowell:70} the thermal conductivity $\kappa(H)$ 
was measured in both monotonically increasing and decreasing field. In Fig.~3
the increasing field ZFC measurements of $c_{\rm e}(T, H)$ exhibit an 
upward curvature above $H_{\rm en}$. This indicates that the flux has some 
difficulty entering the sample even above $H_{\rm en}$.
We note that the measurements of Ref.~\cite{Lowell:70} are often cited in 
the literature as the most archetypal
example of $\kappa(H)$ for a clean conventional type-II superconductor.
However, here we see that the intervortex transfer of quasiparticles at
low fields in Nb is greatly enhanced for a highly-ordered vortex lattice 
generated under FC conditions.

The crossover field for an isotropic-gapped superconductor was determined 
in Ref.~\cite{Nakai:04} to be $B^*(0.1$~$T_c) \! \approx \!  0.33 H_{c2}$.
Here we find for Nb that $B^*(0.1$~$T_c) \! \leq \! H_{\rm en}(0.1$~$T_c)
\! \approx \! 0.21  H_{c2}$. The ratio of the minimum to maximum 
superconducting energy gaps on different Fermi sheets in Nb is estimated 
to be $\sim \! 0.8$ \cite{Crabtree:87}. According to Ref.~\cite{Nakai:04}, 
gap anisotropy of this size reduces the crossover field to 
$B^*(0.1$~$T_c) \! \approx \!  0.25 H_{c2}$. Thus, the small difference 
in gap values may be sufficient to explain the complete absence of a crossover 
field in the vortex state of Nb.       
    
In summary, the $T$-linear coefficient of the electronic specific heat 
$\gamma(T, B)$ has been determined for pure Nb in the vortex state. 
Contrary to popular belief, $\gamma(T, B)$ is not a linear function of $B$
at fields immediately above $H_{c1}$. Our results support
theoretical calculations showing that even in a simple BCS type-II superconductor, 
$\gamma(T, B)$ is a sublinear function of $B$ over a wide region of the vortex phase. 
This calls into question specific heat studies on type-II 
superconductors that have reported $\gamma(B) \! \propto \! B$ 
behavior persisting up to fields close to $H_{c2}$. 
We suspect inaccuracies in the $T \! \rightarrow \! 0$~K
extrapolation procedure are partially responsible, 
but also stress the importance of small field increments in specific heat 
measurements of this kind. 
Lastly, the effect of delocalized quasiparticle core states on the field
dependence of the specific heat should be considered when attempting 
to identify the pairing symmetry of a newly discovered superconductor by 
this method. In other words, the absence of $\gamma(B) \! \propto \! B$
behavior is not necessarily an indication of unconventional superconductivity.

We thank K.~Machida for fruitful discussions. 
The work presented here was supported by the Natural Science and Engineering Research
Council of Canada, and the Canadian Institute for Advanced Research. Work at
Los Alamos was performed under the auspices of the US Department of Energy.
 

\begin{thebibliography}{12} 
 
\bibitem{Caroli:64} C.~Caroli, P.G.~de Gennes and J.~Matricon,
Phys.~Lett. {\bf 9}, 307 (1964).

\bibitem{Fetter:69} A.L.~Fetter and P.~Hohenberg, in {\it Superconductivity},
edited by R.D.~Parks (Marcel Dekker, Inc., New York, 1969),
Vol. 2, pp. 817-923.

\bibitem{Volovik:93} G.E.~Volovik, JETP Lett. {\bf 58}, 469 (1993).

\bibitem{Kramer:74} L.~Kramer and W.~Pesch, Z.~Phys {\bf 269}, 59 (1974); W.~Pesch
and L.~Kramer, J.~Low~Temp.~Phys. {\bf 15}, 367 (1974).

\bibitem{Boaknin:03} E.~Boaknin {\it et al.}, Phys.~Rev.~Lett. {\bf 90},
117003 (2003).

\bibitem{Sonier:04} J.E.~Sonier {\it et al.}, Phys.~Rev.~Lett. {\bf 93},
017002 (2004).

\bibitem{Nakai:04} N.~Nakai, P.~Miranovi\'{c}, M.~Ichioka, and K.~Machida,
Phys.~Rev.~B {\bf 70}, 100503(R) (2004).

\bibitem{Sanchez:95} D.~Sanchez {\it et al.}, Physica~B {\bf 204},
167 (1995).

\bibitem{Sonier:99} J.E.~Sonier, M.F.~Hundley, J.D.~Thompson, and
J.W.~Brill, Phys.~Rev.~Lett. {\bf 82}, 4914 (1999).

\bibitem{Hanaguri:03} T.~Hanaguri {\it et al.}, Physica~B {\bf 329-333},
1355 (2003).

\bibitem{Yokoya:01} T.~Yokoya {\it et al.}, Science {\bf 294}, 2518 (2001). 

\bibitem{Nakai:02} N.~Nakai, M.~Ichioka and K.~Machida, J.~Phys.~Soc.~Jpn. {\bf 71}, 23 (2002);
M.~Ichioka, K.~Machida,  N.~Nakai and P.~Miranovi\'{c}, Phys. Rev. B 70, 144508 (2004).

\bibitem{Callaghan:05} F.D.~Callaghan, M.~Laulajainen, C.V.~Kaiser, and
J.E.~Sonier, Phys.~Rev.~Lett. {\bf 95}, 197001 (2005).

\bibitem{Sosolik:03} C.E.~Sosolik {\it et al.}, Phys.~Rev.~B {\bf 68},
140503(R) (2003).

\bibitem{Yethiraj:99} M.~Yethiraj, D.K.~Christen, D.~McK. Paul,
P.~Miranovi\'{c}, and J.R.~Thompson, Phys.~Rev.~Lett. {\bf 82}, 5112 (1999).

\bibitem{Sonier:04b} J.E.~Sonier, J.~Phys.: Condens. Matter {\bf 16}, S4499 (2004). 

\bibitem{Forgan:02} E.M.~Forgan {\it et al.}, Phys.~Rev.~Lett. {\bf 88}, 167003 (2002).

\bibitem{Bean:64} C.P.~Bean and J.D.~Livingston,  Phys.~Rev.~Lett. {\bf 12}, 14 (1964).

\bibitem{Ichioka:99} M.~Ichioka, A.~Hasegawa, and K.~Machida, Phys.~Rev.~B {\bf 59},
184 (1999). 

\bibitem{Lowell:70} J.~Lowell and J.B.~Sousa, J.~Low.~Temp.~Phys. {\bf 3}, 65 (1970).

\bibitem{Crabtree:87} G.W.~Crabtree, D.H.~Dye, D.P.~Karim, S.A.~Campbell,
and J.B.~Ketterson, Phys.~Rev.~B {\bf 35}, 1728 (1987).

\end{thebibliography}
\end{document}